\documentclass[reprint,amsmath,amssymb, aps]{revtex4-2}

\usepackage{graphicx}
\usepackage{dcolumn}
\usepackage{bm}
\usepackage{soul}
\setstcolor{red}
\usepackage{physics}
\usepackage{xcolor}
\usepackage{hyperref}

\begin{document}

\preprint{APS/123-QED}

\title{Turbulent dynamics in two-dimensional paraxial fluid of light}

\author{Myrann Baker-Rasooli}
\affiliation{Laboratoire Kastler Brossel, Sorbonne University, CNRS, ENS-PSL University, Coll\`ege de France; 4 Place Jussieu, 75005 Paris, France}

\author{Wei Liu}
\affiliation{Laboratoire Kastler Brossel, Sorbonne University, CNRS, ENS-PSL University, Coll\`ege de France; 4 Place Jussieu, 75005 Paris, France}

\author{Tangui Aladjidi}
\affiliation{Laboratoire Kastler Brossel, Sorbonne University, CNRS, ENS-PSL University, Coll\`ege de France; 4 Place Jussieu, 75005 Paris, France}

\author{Alberto Bramati}
\affiliation{Laboratoire Kastler Brossel, Sorbonne University, CNRS, ENS-PSL University, Coll\`ege de France; 4 Place Jussieu, 75005 Paris, France}

\author{Quentin Glorieux}
\email{quentin.glorieux@lkb.upmc.fr}
\affiliation{Laboratoire Kastler Brossel, Sorbonne University, CNRS, ENS-PSL University, Coll\`ege de France; 4 Place Jussieu, 75005 Paris, France}

\date{\today}

\begin{abstract}
Turbulence in quantum fluids has, surprisingly, a lot in common with its classical counterpart.
Recently, cold atomic gases has emerged as a well controlled experimental platform to study turbulent dynamics.
In this work, we introduce a novel system to study quantum turbulence in optics, with the major advantage of having access to a wide range of characterization tools available for light fields.
In particular we report the temporal dynamics of density and phase and we show the emergence of isotropy in momentum space and the presence of different scaling laws in the incompressible kinetic energy spectrum.
The microscopic origin of the algebraic exponents in the energy spectrum is discussed by studying the internal structure of quantized vortices within the healing length and their clustering at larger length scales.
These results are obtained using two counter-streaming fluids of light, which allows for a precise preparation of the initial state and the in-situ measurement of the compressible and incompressible fluid velocity.
 \end{abstract}

\maketitle

\section{Introduction}
In a pioneering theoretical work, Lars Onsager made the connection between turbulence and point vortex dynamics in a two-dimensional (2D) classical fluid  \cite{onsager1949statistical}.
This model found a direct application for incompressible superfluids such as HeII, where vortices are quantized and can be considered as point-like objects \cite{Niemela2005}.
In consequence, the study of quantum turbulence \cite{Vinen2002}, i.e. the chaotic motion in quantum fluids such as superfluids has largely been motivated by the expectation that understanding the dynamics of vortices in quantum fluids will help our comprehension of the general nature of turbulence.
However, due to the strong interactions, a quantitative microscopic model does not exist for superfluid helium which complicate the comparison between theory and experiments \cite{Nore1997} and ultra-cold atoms have emerged as a platform of choice to study quantum turbulence \cite{nazarenkoWaveTurbulenceVortices2006, Tsubota2009b, white2014vortices, tsatsos2016quantum}.
Atomic BECs have allowed to observe Kelvin waves cascade \cite{kivotides2001kelvin}
and vortex tangle in superfluid turbulence \cite{PhysRevLett.103.045301}.
More recently, 2D turbulence have been studied in BEC and important phenomena have been evidenced including vortex clustering \cite{Gauthier2019}, direct energy cascade \cite{Navon2019}, bidirectional dynamic scaling \cite{Glidden2021} and self-similarity dynamic \cite{galkaEmergenceIsotropyDynamic2022}. 

\begin{figure*}[ht!]
    \centering
    \includegraphics[width=0.97\linewidth]{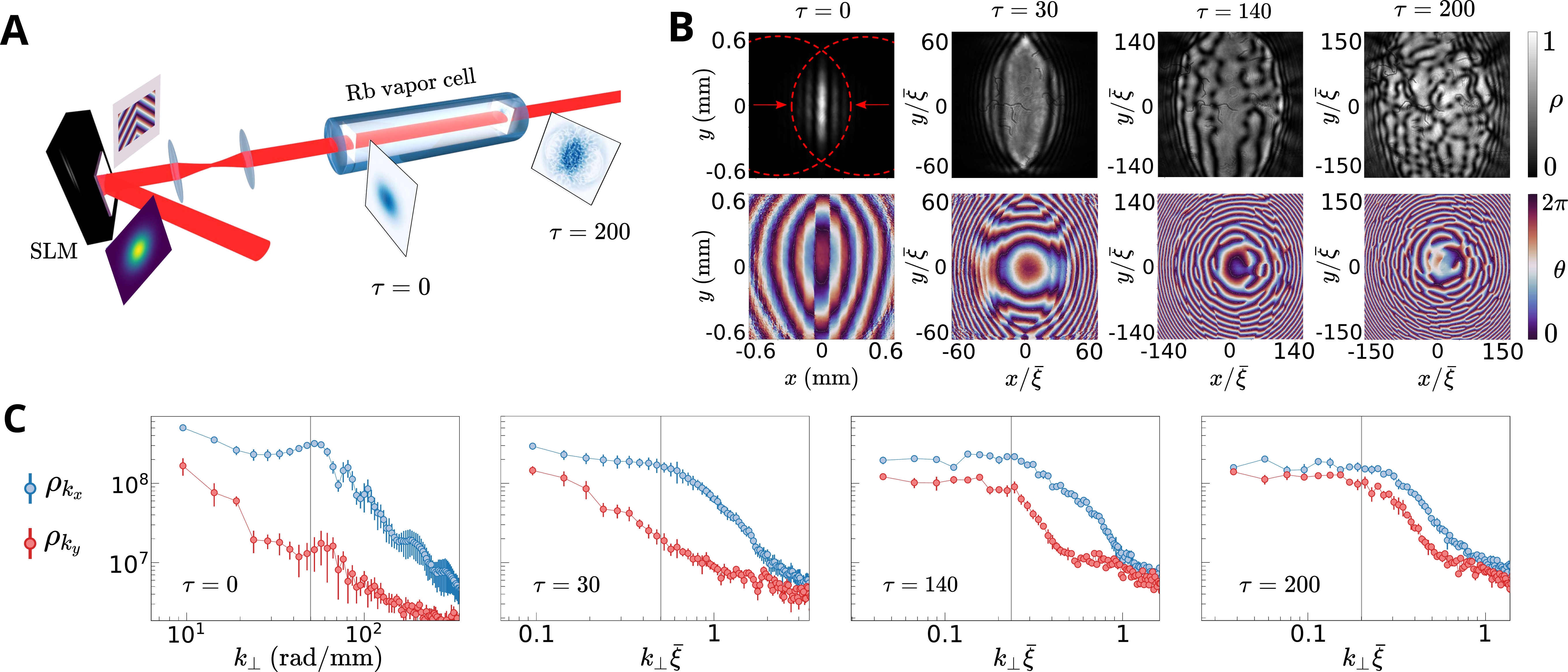}
    \caption{\textbf{Emergence of isotropy.} \textbf{A} - Simplified setup. A 780 nm laser beam is sent on a SLM imaged at the input of the 5~cm-long Rb vapor cell. The phase modulation leads to two counter-propagating beams in the transverse plane. The output plane of the non-linear medium is imaged on a camera.
     \textbf{B} - Top: Experimental images of the intensity. Bottom: Associated phase. From left to right $\tau=0,30,140,200$.
     \textbf{C} - Averaged density spectrum along $k_x$ (blue) and $k_y$ (red) over 20 random realizations for different values of $\tau$. The solid gray line represents the forcing frequency ($k_f = 2\pi\cdot 8~\text{mm}^{-1}$), imposed by the angle between the two fluid set on the SLM. For all $\tau>0$, $k_f\Bar{\xi}$ is kept constant and $\Bar{\xi}=4~\mu m$ at $\tau=200$.}
    \label{fig:setup}
\end{figure*}

In this work, we introduce a novel platform for studying quantum turbulence based on paraxial fluids of light in hot atomic vapors.
In this system, superfluidity in 2-dimensions has been demonstrated \cite{Fontaine2018,Fontaine2020} and several hydrodynamical effects (including shockwaves and blast waves) have been observed experimentally \cite{abuzarli2021blast,Azam2021a,Bienaime2021,PhysRevA.105.043510}.
Thanks to optical wavefront-shaping and optical detection techniques this non-linear system gives access to novel observables for quantum turbulence research such as the velocity field and the spectra associated with well controlled few-vortex configurations.
In this letter, we present how to exploit these optical techniques to temporally resolve the 2D dynamics of  turbulence in a quantum fluid of light.

The turbulence is initiated by colliding two counter-streaming fluids of light with controlled velocities and tunable densities \cite{rodrigues2020turbulence} and we directly measure the fluid kinetic energy, which we decompose into compressible (linked to phononic-excitations) and incompressible (linked to vortices) parts.
We first describe the emergence of isotropy in the system, even though we force it along a preferential direction.
We then study the spontaneous appearance of vortices and their clustering; with their role in the appearance of power law in the kinetic energy spectrum at known length scales. 
As expected, we observe that the internal structure of the vortices, within the healing length, is responsible for the scaling exponent in the large momentum limit.
Finally, we demonstrate that the single vortex and vortex pair spectrum follows universal exponent as predicted in \cite{Bradley2012}.

\section{Paraxial fluid of light and emergence of isotropy}
\subsection{Paraxial fluid of light}
A paraxial fluid of light consists of a monochromatic laser beam propagating through a non-linear medium. 
In the paraxial approximation, the propagation equation of the laser electric field envelope $\mathcal{E}$ is isomorphic to the Gross-Pitaevskii equation describing the temporal evolution of the wavefunction for a weakly interacting quantum gas \cite{carusotto2014superfluid}.
This reads as: 
\begin{equation}
    \label{eq:NLSE}
    i \pdv{\mathcal{E}}{z} \ (\mathbf{r}_\perp, z ) =  \left[ -\frac{1}{2 n_0 k_0} \nabla_\perp^2 - \frac{k_0 \chi^{(3)}}{2 n_0} |\mathcal{E} (\mathbf{r}_\perp, z )|^2  \right]\mathcal{E} (\mathbf{r}_\perp, z )~.
\end{equation}
where  $k_0$ is the wavevector, $n_0$ is the linear refractive index given by $n_0 = \sqrt{1 + \text{Re}\chi^{(1)}}$, the nabla operator $\nabla_\perp$ must be understood as acting in the transverse $\mathbf{r}_\perp=(x,y)$ plane as a consequence of the paraxial approximation and the non-linear term is proportional to the third-order susceptibility at the laser frequency $\chi^{(3)}$ times the laser intensity.
This last term induces an effective photon-photon interaction, and is set negative to ensure a stable superfluid with repulsive interactions.
In this formalism, each transverse plane at fixed z, is a temporal snapshot of the evolution for an ideal 2D system.
\subsection{Effective time and adimensional equation}
The light field within the non-linear medium is not directly accessible experimentally and the temporal evolution.
We can adimension the propagation equation of the laser electric field by incorporating the fluid interaction into a new variable $\tau=~{L}/{z_{\text{NL}}}$, 
where  $z_{\text{NL}}=\left[-\frac{k_0}{2 n_0} \chi^{(3)} |\mathcal{E} (\mathbf{r}_\perp, z )|^2\right]^{-1}$ is the characteristic non-linear axial length and $L$ is the length of the non-linear medium.
After re-scaling the transverse quantities ($\tilde{\textbf{r}}=\textbf{r}/\Bar{\xi}$, $\tilde\nabla_{\perp} = \Bar{\xi}\nabla_{\perp}$) by the transverse healing length $\Bar{\xi}= \sqrt{\frac{z_{\text{NL}}}{k_0}}$, one obtains for $\psi=\frac{\mathcal{E}}{|\mathcal{E}|}$:
\begin{equation}
    i\frac{\partial\psi }{\partial \tau}=
    \left(-\frac{1}{2}\tilde\nabla^2_{\perp}+{\mid}\psi{\mid}^2
    \right)\psi.
    \label{GPE_Adim}
\end{equation}
In this form, it becomes clear that following the evolution of the system can be done by tuning the ratio $\tau=~{L}/{z_{\text{NL}}}$ and in particular by tuning the non-linear susceptibility $\chi^{(3)}$ \cite{abuzarli2021blast,Bienaime2021}..

\subsection{Experimental description}
In the experiment, we create two fluids of light by propagating a 780 nm diode-laser set close to the resonance of the $^{87}$Rb D2 line within a warm vapor cell of rubidium which acts as a non-linear medium. 
As suggested in \cite{rodrigues2020turbulence} and pictured in Fig.\ref{fig:setup}A , we inject two counter-streaming fluids of light by imposing opposite wavevectors to the right and the left side of the laser beam with a Spatial Light Modulator (SLM). 
The angle between the wavevectors on the two sides imposes a relative velocity to the fluids and defines the forcing spatial frequency $k_f$.
Experimentally, we adjust $k_f$ between $2\pi\cdot 8\textrm{ mm}^{-1}$ and $2\pi\cdot 45\textrm{ mm}^{-1}$ in order to keep the quantity $k_f\Bar{\xi}$ constant while changing $\tau$, with $\Bar{\xi}$ being the mean healing length  accounting for the losses during the propagation (see Appendix A).

The intensity and phase are recorded at the output of the cell ($L=5$~cm) and typical images are shown in Fig.\ref{fig:setup}B (top: intensity, bottom: phase) for an increasing effective time $\tau$.
The initial state ($\tau=0$) corresponds to the linear regime and is obtained by setting a large detuning ($-5$~GHz) and low laser power ($P=10$~mW).
Longer effective times ($\tau=30,140,200$) are obtained by setting the detuning closer to resonance and a higher power, therefore reducing $z_{\text{NL}}$.
Observation of the density images in Fig.\ref{fig:setup}B reveals the presence of instabilities which develops into pairs of vortices and anti-vortices with increasing $\tau$.
In this geometry with counter-streaming flows, we force the system along the $x$-direction at short length-scale, and we will first verify that turbulence develops as an anisotropic phenomenon across the system \cite{falkovich1993revised,galkaEmergenceIsotropyDynamic2022}.

\subsection{Isotropy emergence from an anisotropic forcing}
In Fig.\ref{fig:setup}C, we present the density spectrum along $k_x$ (blue) and $k_y$ (red) obtained by fast Fourier transform of the density images, for different propagation times and averaged over 20 realizations.
We calibrate our spatial resolution in order to ensure that we resolve the inner structure of the vortices ($<14\mu m$) (see Apendix B).
For the initial state ($\tau=0$), a clear modulation at the forcing length-scale   $k_f=2\pi\cdot 8\textrm{mm}^{-1}$ is visible along $k_x$ due to the angle imposed between the two fluids by the SLM modulation.
This initial state reflects the anisotropic forcing at short length-scale.
We observe a clear reduction of the forcing peak with the increase of $\tau$ increases, but at short effective time ($\tau\leq140$) the density spectrum remains anisotropic.
Finally, the system evolves toward an isotropic configuration for $\tau=200$, for which $\rho_{k_x}$ and $\rho_{k_y}$ overlap at all $k$.
This observation demonstrates the emergence of statistical isotropy under anisotropic forcing, in agreement with a recent observation in ultra-cold atomic gases \cite{galkaEmergenceIsotropyDynamic2022}.\\

\section{Velocity decomposition and vortex clustering}
\subsection{Velocity decomposition}
\begin{figure}[]
    \centering
    \includegraphics[width=1\linewidth]{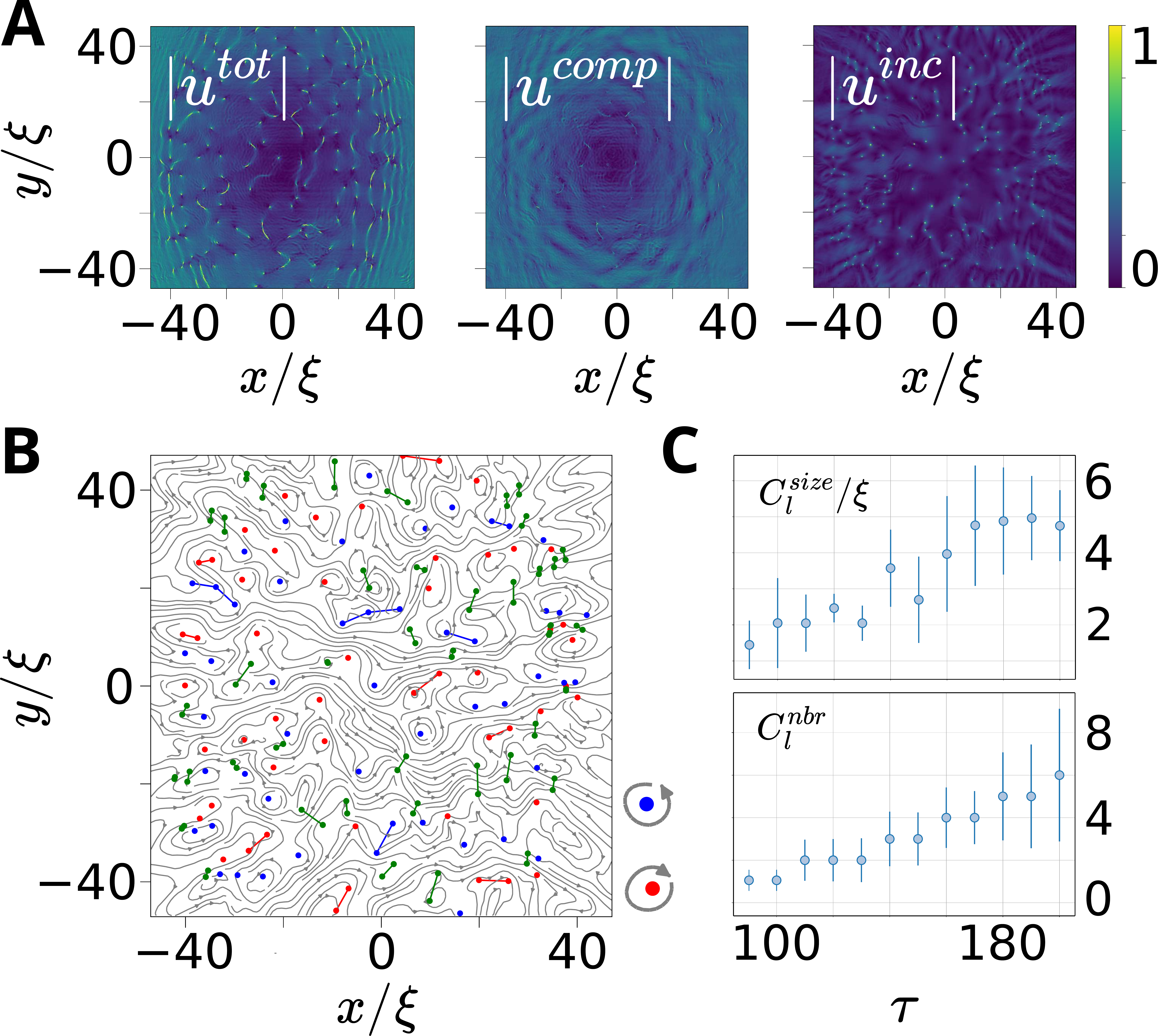}
    \caption{\textbf{Velocity decomposition and vortex clustering.} \textbf{A} - Modulus of each velocity field components. 
    \textbf{B} Streamlines calculated from $\boldsymbol{u}^{\textbf{inc}}(\textbf{r})$.
    The red and blue dots are the position of the positive and negative vortices linked by a solid line, blue or red if they belong to the same cluster, green if they belong to a dipole.
    \textbf{C} -  Average cluster number $C_l^{nbr}$ and cluster size $C_l^{size}$ in $\xi$ unit}
    \label{fig:velocity}
\end{figure}
In contrast with standard cold-atoms experiments, the use of fluid of light allows for a direct access to the fluid's phase as shown in Fig.\ref{fig:setup}B. 
This allows us to measure the velocity field, given by $\boldsymbol{v^{tot}}(\textbf{r}) \propto \nabla_{\perp}\theta(\textbf{r})$ (see Apendix E).
To compute the kinetic energy we introduce the density-weighted velocity, given by $\boldsymbol{u^{tot}(r)}=\sqrt{\rho(\boldsymbol{r}})\boldsymbol{v}^{tot}(r)$, where $\rho(\boldsymbol{r})$ is the light intensity.
A typical example is shown in Fig.\ref{fig:velocity}A (left) at $\tau=200$. 
We then identify the compressible and incompressible parts of $\boldsymbol{u^{tot}(r)}$ using the Helmholtz decomposition  \cite{horng2009two,panico2022onset} to separate the divergent (compressible) and rotational (incompressible) components:
\begin{equation}
    \boldsymbol{u^{tot}}(\boldsymbol{r})=\underbrace{\boldsymbol{\nabla}\phi(\textbf{r})}_{\textrm{compressible}} + \underbrace{\boldsymbol{\nabla}\times\textbf{A}(\textbf{r})}_{\textrm{incompressible}}
\end{equation}
where $\phi$ is an arbitrary scalar and $\textbf{A}$ an arbitrary vector field. 
The same decomposition can be written in the Fourier space:
\begin{equation}
    \boldsymbol{U^{tot}}(\boldsymbol{k}) = i\boldsymbol{k}U_\phi(\boldsymbol{k})+i\boldsymbol{k}\times\boldsymbol{U_{A}}(\boldsymbol{k}),
\end{equation}
where $U_\phi(\boldsymbol{k})=-i\frac{\boldsymbol{k}\cdot \boldsymbol{U^{tot}}(\boldsymbol{k})}{||\boldsymbol{k}||^2}$ and $\boldsymbol{U_{A}}(\boldsymbol{k})=i\frac{\boldsymbol{k}\times \boldsymbol{U^{tot}}(\boldsymbol{k})}{||\boldsymbol{k}||^2}$.\\
Thus, we  write the definition of the compressible and incompressible part in the real space:
\begin{equation}
    \begin{split}
        \boldsymbol{\nabla}\phi(\boldsymbol{r})&=\textrm{TF}^{-1}[i\boldsymbol{k}\cdot U_\phi(\boldsymbol{k})] \\
        \boldsymbol{\nabla}\times\textbf{A}(\boldsymbol{r})&=\textrm{TF}^{-1}[i\boldsymbol{k}\times \boldsymbol{U_A}(\boldsymbol{k})].
    \end{split}
\end{equation}

\begin{figure*}[ht!]
    \centering
    \includegraphics[width=1\linewidth]{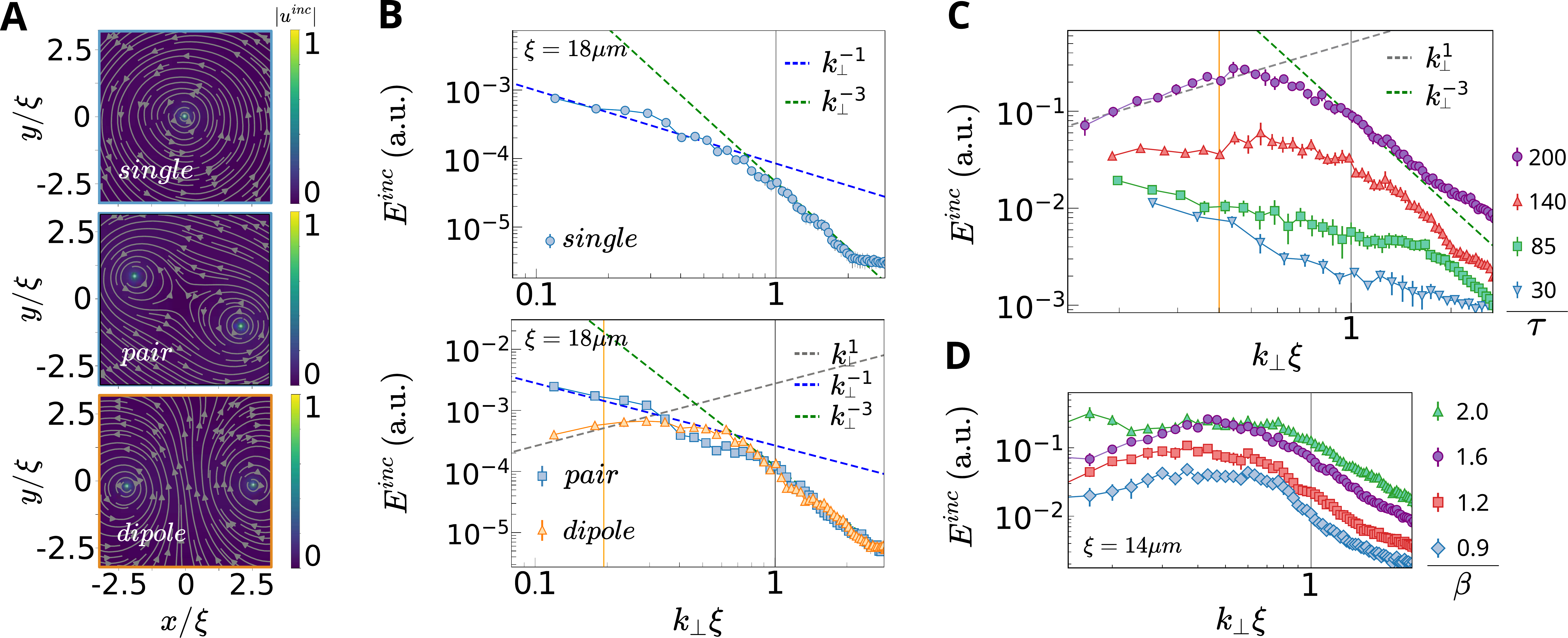}
    \caption{\textbf{Incompressible kinetic energy spectrum.} 
    \textbf{A} - Streamlines and their associated $|u^{inc}|$ for a single vortex, a vortex dipole (opposite sign) and a vortex pair (same sign), from top to bottom.
    \textbf{B} - Associated energy spectra. Those configuration are injected directly in the fluid using the SLM. 
    Top: blue dots corresponds to a single vortex.
    Bottom: orange triangles corresponds to a dipole (opposite sign) and blue squares to a pair (same sign).
    In both figures, dashed lines are power-law scaling with exponent -1 (blue), -3 (green), +1 (gray).
    \textbf{C} - Energy spectrum for N vortices, generated with the two counter-streaming fluids, for a Mach number $\beta=v/c_s=1.6$ and $\tau=30,85 140, 200$. 
    The solid orange line represents the lower bound of the emerging range set at the inverse of the mean cluster radius, $1/(35\mu m)$, 
    and the vertical gray line is the upper bound at $1$ (fixed by the healing length at the output which is typically $14\mu m$ for $\tau=200$).
    The grey and green dashed line are power-law function ploted respectively for $k^1$ and $k^{-3}$. 
    \textbf{D}: Spectra for different Mach numbers $\beta=0.9,1.2,1.6,2.0$ at fixed time $\tau=200$.
   }
    \label{fig:spectra}
\end{figure*}

The incompressible velocity is calculated by subtracting the compressible part from the total velocity $\boldsymbol{u^{inc}}=\boldsymbol{u^{tot}}- \boldsymbol{\nabla}\phi(\textbf{r})$ shown in Fig.\ref{fig:velocity}A (right).
The associated streamlines are displayed in Fig.\ref{fig:velocity}B and overlapped with vortex and anti-vortex which are identified by the winding direction of their phase and are plotted in red and blue.

\subsection{Vortex clustering}
Using a nearest neighbor vortex classification from \cite{Reeves2013}, we sort vortices into opposite sign dipoles (green) and same sign clusters represented by the solid links in Fig.\ref{fig:velocity}B.
The classification is then used to give quantitative elements on the clustering dynamics.
Fig.\ref{fig:velocity}C we report the averaged cluster number $C_l^{nbr}$ (bottom) and normalized cluster size $C_l^{size}/\xi$ versus the propagation time $\tau$ over 20 random realizations.
At short time ($\tau<90$), no clustering mechanism is observed and only dipoles remains.
By increasing $\tau$, clustering start occurring and $C_l^{nbr}$ grows progressively with $C_l^{size}/\xi$, indicating larger scales appearing in the system.
The large uncertainty is a consequence of the multiple realizations, showing that the vortex distribution strongly depends on the initial conditions.\\

\section{Incompressible energy spectra}
\subsection{Kinetic energy computation}
With the incompressible density-weighted velocity field, we compute the incompressible kinetic energy spectrum.
The total kinetic energy is written as $E_{kin}=\frac{m}{2} \iint \rho(\boldsymbol{r})|\boldsymbol{v^{tot}}(\boldsymbol{r})|^2\textrm{d}x\textrm{d}y$, with the photon mass $m=\frac{\hbar k_0}{c}\sim10^{-36}\textrm{ kg}$. In the Fourier space its incompressible part can be written as:
\begin{equation}
    E_{kin}^{inc}=\frac{m}{2}\iint\left(|U_x^{inc}(\boldsymbol{k})|^2 + |U_y^{inc}(\boldsymbol{k})|^2\right)\textrm{d}_x\textrm{d}_y,
\end{equation}
with $\boldsymbol{U^{inc}}_{\boldsymbol{i}}=\textrm{TF}[\boldsymbol{u^{inc}_i}]$. 
Finally, in the turbulent regime our system tends towards a quasi-isotropic behavior. 
We consider the longitudinal spectrum of the incompressible kinetic energy, obtained by integrating over the azimuthal angle as :

\begin{equation}
    E_{kin}^{inc}=\frac{m}{2}k\int^{2\pi}_0\left(|U_x^{inc}(\boldsymbol{k})|^2 + |U_y^{inc}(\boldsymbol{k})|^2\right)\textrm{d}\Omega_k.
\end{equation}

Thanks to the use of a SLM it is possible to shape the initial state with well defined initial condition in order to measure the experimental incompressible kinetic energy spectra as presented in Fig.\ref{fig:spectra}.

\subsection{One and two vortex configurations}
We first generate three simple configurations by direct imprinting with the SLM: a single vortex, a pair (same sign) and a dipole (opposite sign) and we present the experimental density and streamlines of the velocity field in Fig.\ref{fig:spectra}A.
The experimental incompressible kinetic energy spectra are shown in Fig.\ref{fig:spectra}B. 
For a state containing only a single-vortex (and therefore no acoustic energy), the field is automatically incompressible \cite{Bradley2012}.
The spectrum of a single vortex state shows a $k^{-1}$ decay in the infrared (IR) range ($k_\perp \xi \ll 1$) which arises purely from the irrotational velocity field of a quantum vortex. 
The scaling is similar to point-like vortices and is the only remaining signature of the vortex far from its core.
However, in contrast with the point-like model, in the ultraviolet (UV) range ($k_\perp \xi \gg 1$) we observe a $k^{-3}$ decay which stems from the internal structure of the vortex core as described theoretically in \cite{PhysRevLett.105.129401,Bradley2012}.

As expected, the UV scaling remains identical for the pair and the dipole spectra, since the vortex core structure is not modified, but the additional kinetic energy of the two-vortices state is observed throughout the spectrum in Fig.\ref{fig:spectra}B as a vertical shift by a factor close to 2 (the exact ratio as function of $k$ is given in the Supplementary material). 
Moreover the IR behavior differs from the isolated case. 
For the pair spectrum, we find a $k^{-1}$ IR-scaling similar to the single vortex configuration, which arises from the fact that we cannot distinguish the velocity field of a single vortex from a pair when we observe far from the structure, as suggested by Fig.\ref{fig:spectra}A.
On the contrary, a $k^{1}$ IR-scaling is observed for the dipole configuration which originates from the cancellation of the velocity field for length scales much larger than typical inter-vortex separation in any neutral configuration of vortices \cite{Bradley2012}.

\begin{figure}[hb!]
    \centering
    \includegraphics[width=0.97\columnwidth]{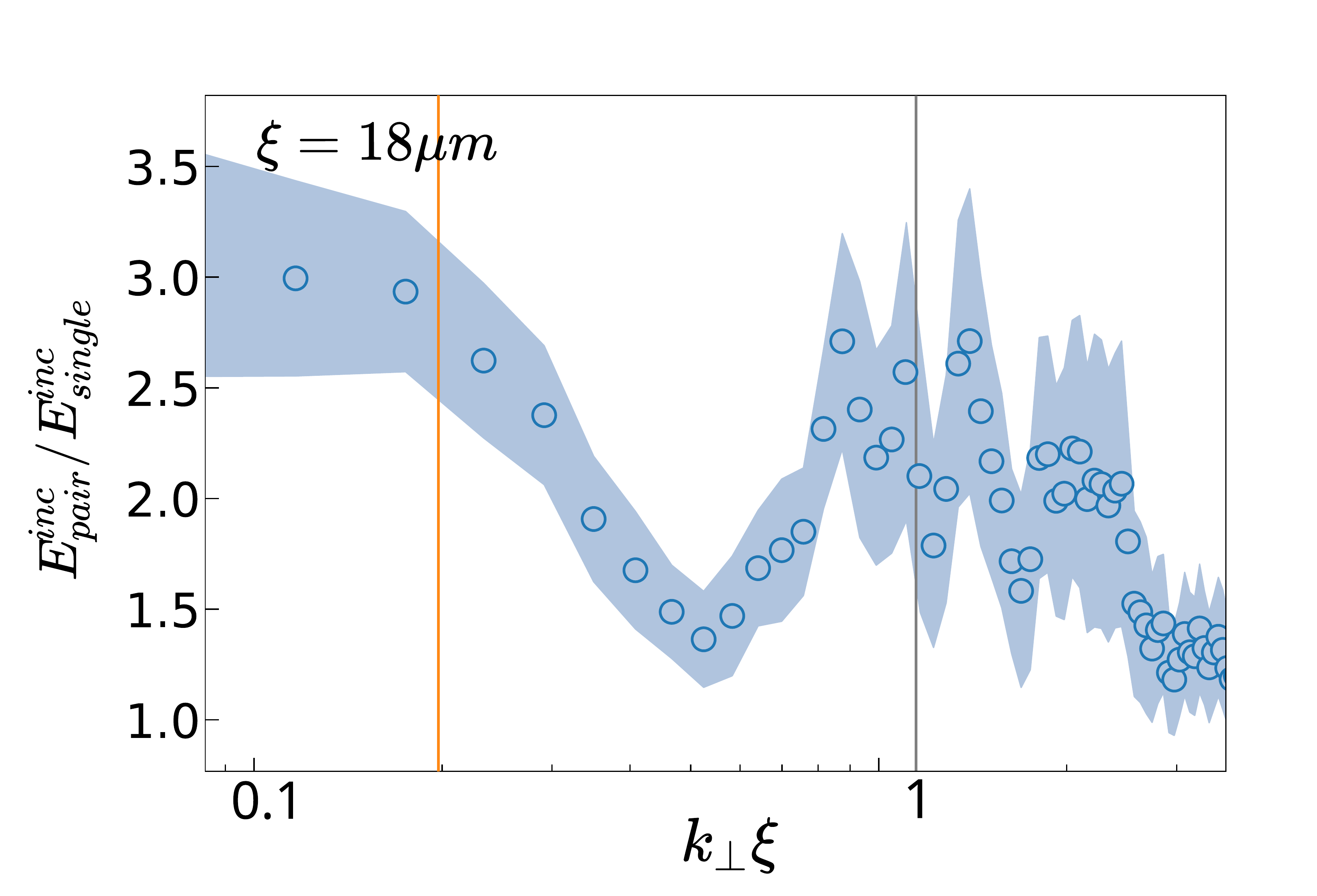}
    \caption{\textbf{Ratio of a pair of vortices with a single vortex} - 
    A average factor of $2.0\pm0.3$ is observed in the inter-vortex region.
    The orange line represents the inverse of the inter-vortex distance at 1/(90µm), and the vertical gray line is the upper bound at 1 (fixed by the healing length at the output plane).}
    \label{fig:ratio}
\end{figure}

In order to compare the two-vortex configuration with the single-vortex one, we calculate the ratio of the kinetic energy spectra between a pair and a single vortex as shown Fig.\ref{fig:ratio}.
An average factor of $2.0\pm0.3$ is observed in the region between the inter-vortices distance and the healing length $\xi$. 
Moreover, we observe that at distances below the size of the vortices this ratio tends towards 1, which is expected since it does not depend on the number of vortices in the system.
\subsection{N-vortex configuration}
Finally, we present the experimental kinetic energy spectra in the case of two-counter propagating superfluids of light in Fig.\ref{fig:spectra}C for different propagation time $\tau$.
Interestingly, we find the same $k^{-3}$ UV-scaling (due to the vortex cores) and the same $k^{1}$ IR-scaling observed for the dipole configuration (since it is also a neutral configuration).
In between these two regions, a spectral range with an intermediate scaling emerges as function of $\tau$. 
This range is absent at short time ($\tau=30$) and emerges progressively with $\tau$ from the scale of the forcing frequency towards the IR.
For $\tau=200$, a fit of this intermediate range gives a $k^{-5/3}$ scaling, however the frequency range is too narrow to conclude on the physical origin of this exponent \cite{Numasato2009,Bradley2012,Reeves2013,Shukla_2013,zhu2022direct,PhysRevA.86.053621,PhysRevLett.113.165302}, since, for example, accidental scaling laws have been predicted in \cite{PhysRevA.91.023615}.
An important point is that the low frequency cut-off of this intermediate range is given by the average cluster size as plotted in Fig. \ref{fig:velocity}C. 
Since this cut-off still evolves to lower frequency with $\tau$ (for $\tau=200$), this confirms that our system has not reached the steady state yet and that a longer evolution time would allow for a broader spectral range, with potentially several decades of width.

Finally, we took advantage of the optical setup to tune the relative velocity of the counter-streaming fluids using the SLM and report it in Fig.\ref{fig:spectra}D.
By tuning the fluid velocities at fixed $\tau$ for Mach numbers $\beta$ (defined as the ratio between the velocity and the speed of sound) between 0.9 and 2.0, we therefore changed the forcing frequency and we verified the presence of a scaling different from $-1$ or $-3$ in the intermediate range for all these Mach numbers.\\

\section{Conclusion}
In this work, we have evidenced and characterized turbulence in a 2D paraxial fluid of light using an anisotropic forcing. 
Thanks to this novel platform, we were able to study the phenomenon during its effective temporal evolution and observe the emergence of isotropy and scaling laws in the energy spectrum.
Moreover, thanks to precise phase measurement, we had access to the velocity field and the vortex distribution and thus study the growing clustering mechanism.
By extracting the incompressible density weighed velocity, using the Helmholtz decomposition, we reported the incompressible kinetic energy spectra in several configurations.
We explained different spectral regions of 1, 2 and N vortices configurations and explored their behavior in time and for different energy injection.
These results introduce a novel platform for studying quantum turbulence and open the way to explore acoustic waves turbulence \cite{PhysRevA.102.043318,PhysRevE.106.014205,PhysRevLett.128.224501} and more generally out-of-equilibrium physics of quantum fluids.

\section*{Acknowledgements}
The authors warmly thank S. Nazarenko, H. Tercas, D. Ballarini, D. Sanvitto, L. Canet, A. Minguzzi, PE Larr\'e, T. Bienaim\'e, T. Joly-Jehenne, M Abuzarli and R. Dubessy for highly valuable discussions on our preliminary experimental results and precise feedbacks on this manuscript.
This work is supported by ANR funding Quantum-SOPHA, the DIM Sirteq, Emergences project from Sorbonne University. Q.G. and A.B. are members of the IUF.

\appendix
\section{Experimental setup}
For creating the fluid of light, we use a $780$nm diode-laser for which we can finely adjust the detuning with respect to the D2 resonance line of $^{87}$Rb. 
A relative phase modulation of wavevector $k_f$ between the right and left side of the beam is imprinted using a Spatial Light Modulator (SLM). 
In order to eliminate the unmodulated reflection on the SLM, we superpose a vertical grating to the horizontal one and select only the first vertical order in the Fourier plane with a slit. 
The SLM is imaged at the entrance of the nonlinear medium with a demagnification to increase the intensity of the beams. 
The beam waist at the medium entrance is $\omega=0.5 \textrm{mm}$.\\
The non-linear medium is a $5$cm rubidium cell containing a mixture of 5$\%$ of $^{87}$Rb and 95$\%$ of $^{85}$Rb. 
The output plane of the cell is imaged on a CMOS camera. Using a reference beam (separate from the main beam upstream) we have access to the phase of our two contra-propagating beams by studying the interference fringes obtained after recombination before the camera.
\begin{figure}[h!]
    \centering
    \includegraphics[width=0.97\columnwidth]{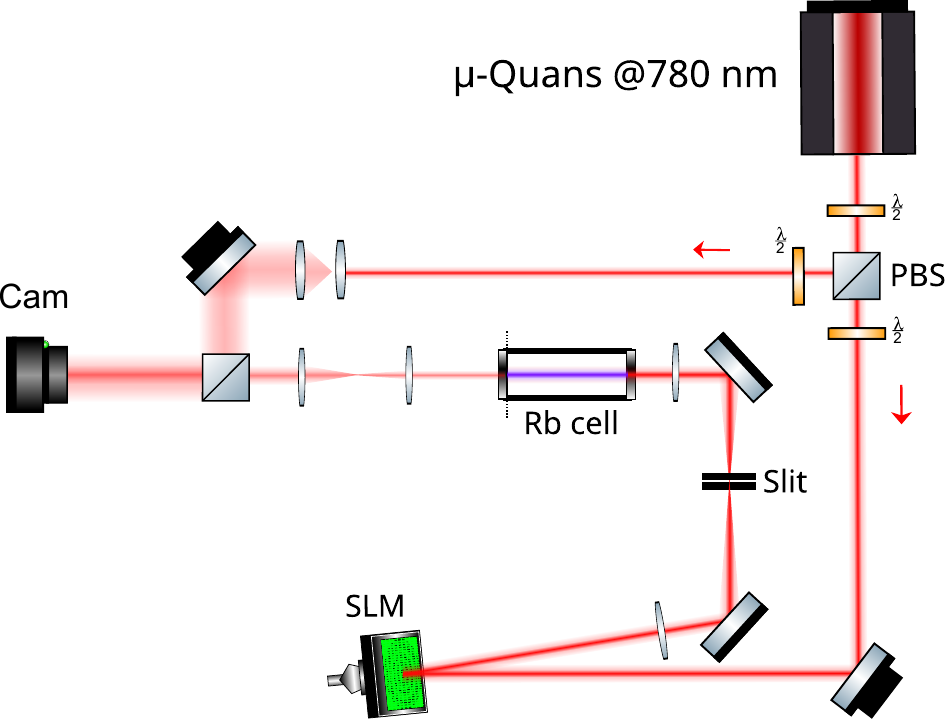}
    \caption{\textbf{Setup detail} - A 780 nm laser beam is sent on a SLM imaged at the input of the 5~cm-long Rb vapor cell. The phase modulation leads to two counter-propagating beams in the transverse plane. The output plane of the non-linear medium is imaged on a camera. A reference beam, which has been separated from the inital laser beam, is recombined with the pump before the camera for our phase measurement.}
    \label{fig:manip_detail}
\end{figure}
The non-interacting case ($\tau=0$) is obtained experimentally by setting a large detuning ($\Delta=-5$GHz) and low laser power (P=10mW).
For the interacting case, the effective time $\tau$ is tuned by adjusting the laser power (up to a maximum of $500$mW) while fixing the cell temperature at 150°C and the laser detuning at $\Delta=-2\textrm{GHz}$ according from the $^{85}$Rb $F = 3$ to $F^{'}$ transition. 

\section{Modulation Transfer Function}
\begin{figure}[h]
    \centering
    \includegraphics[width=0.97\columnwidth]{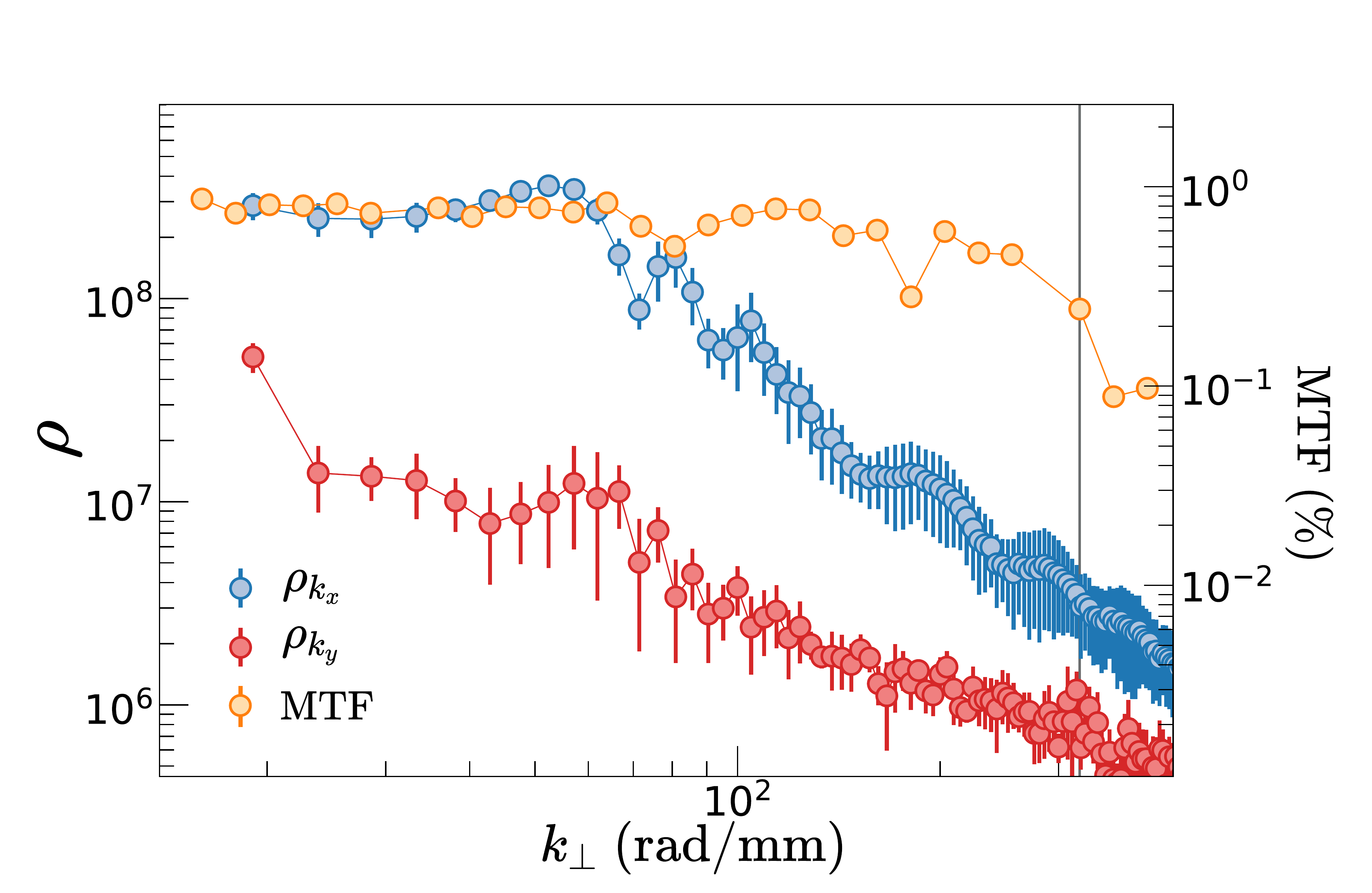}
    \caption{\textbf{MTF} - Modulation Transfer Function for different system syze, overlaped over the density profile along $k_\perp$ (linear case). The 20\% cutoff frequency is represented by a vertical line at $310\textrm{ rad/mm}$.}
    \label{fig:MTF}
\end{figure}
\begin{figure}[hb!]
    \centering
    \includegraphics[width=0.97\columnwidth]{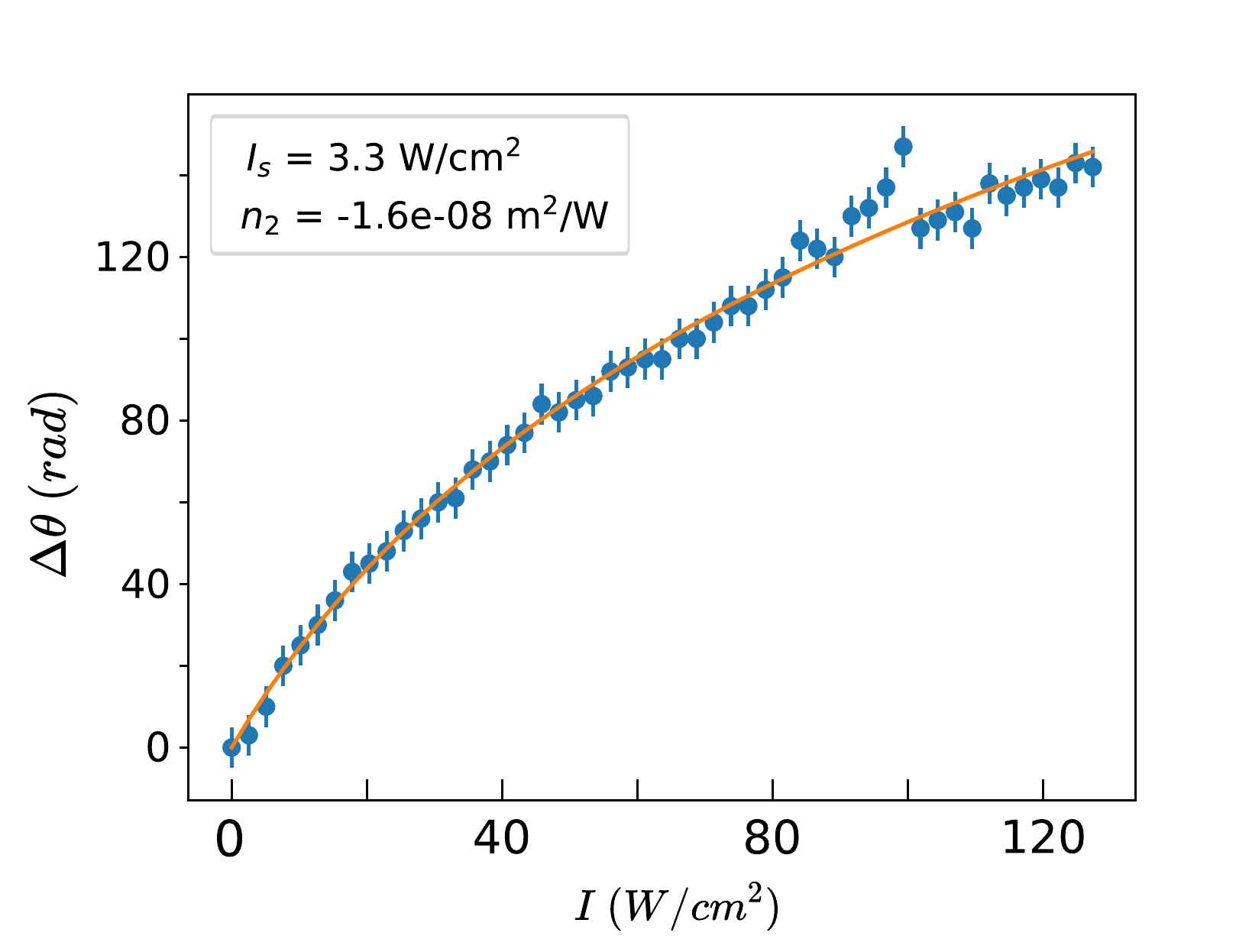}
    \caption{\textbf{Non-linear accumulated phase along the medium} - Each point is measured for a given beam intensity and the data are fitted from (Eq. \ref{fit}), represented by the orange curve.}
    \label{fig:non-linear phase}
\end{figure}
\begin{figure*}[ht!]
    \centering
    \includegraphics[width=0.7\linewidth]{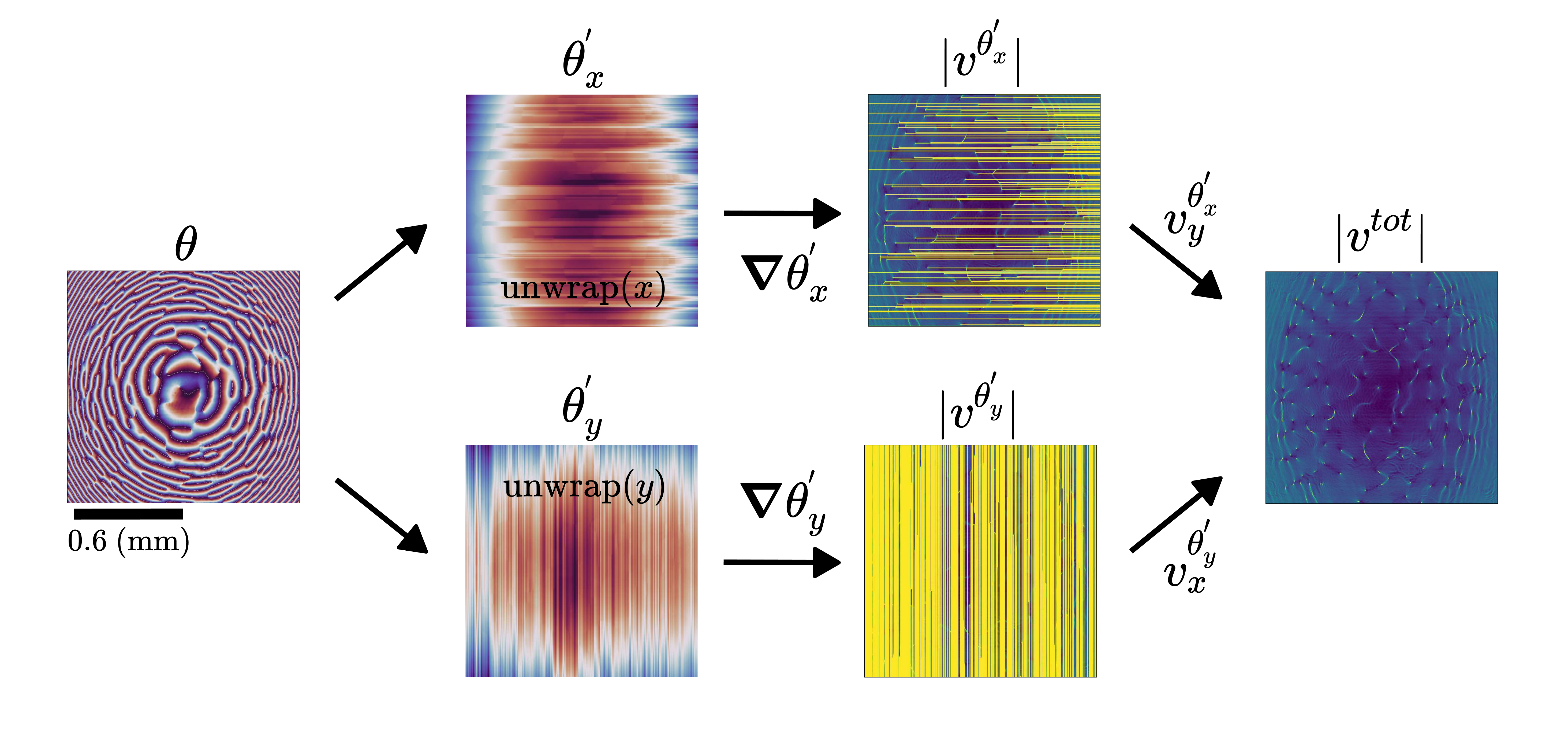}
    \caption{\textbf{Velocity computation process} - The total velocity is the combination of each gradient component calculated from the 1D unwraped phase.}
    \label{fig:velocity_comp}
\end{figure*}
We measure the Modulation Transfer Function (MTF) and we superimpose it on our experimental spectra as shown Fig.\ref{fig:MTF}. The solid line represents the cutoff frequency, showing that our imaging system does not allow us to resolve objects smaller than $3\mu m$.

\section{Influence of losses and quantities}
Our non-linear medium is not entirely described by a lossless pure $\chi^{(3)}$. 
Even though the general effects are not modified, the quantitative measurements provided in the main text requires to take into account loss.

In order to properly measure these quantities in our setup, we let a Gaussian beam propagate in the nonlinear medium and make it interfere with a reference arm on the camera. 
We extract the nonlinear accumulated phase $\Delta \theta$ along the propagation in the Rb cell
This measurement is then performed for different laser intensities (from 0 to 127 $\text{W/cm}^2$), as shown Fig.\ref{fig:non-linear phase}.
We obtain the saturation intensity $I_s$ as well as that of the non-linear index $n_2$ by fitting the data with the solution from:
\begin{equation}
    \label{theta}
    \Delta \theta(I) = k_0 n_2 \int^L_0 \frac{I(z)}{1+\frac{I(z)}{I_s}}.
\end{equation}
Since the intensity in the Rb cell can be written as:
\begin{equation}
    \label{I}
    \frac{\partial I(z)}{\partial z} = -k_0 n_2 \frac{I(z)}{1+\frac{I(z)}{I_s}},
\end{equation}
with $I(z)=I_0 e^{-\alpha z}$, $\alpha$ being the absorption coefficient, and equal to $78\textrm{m}^{-1}$ in our case.
By injecting (\ref{I}) into (\ref{theta}), we get the non-linear phase:
\begin{equation}
    \label{fit}
    \Delta\theta(I) = k_0 n_2 I_s \times \frac{\alpha L - \ln(1/T + I/I_s)}{\alpha},
\end{equation}
with the transmission $T = e^{-\alpha L}$.\\

The non-linear length $z_{\text{NL}}$ depends on the laser intensity and the medium non-linearity.
In our system both quantities are influenced by losses.

In order to take into account these effect and measure $\tau$ and $\Bar{\xi}$ experimentally, we define the averaged non-linear refractive index $\Bar{\Delta n}=n_2 I$ to rewrite the quantities in the form of $z_{NL} = \frac{1}{k_0\Bar{\Delta n}}$ and $\Bar{\xi} = \frac{1}{k_0\sqrt{2\Bar{\Delta n}}}$.

The consequence of the losses is that the measurement of $\Bar{\Delta n}$ does not allows us to know the value of the healing length $\xi$ at the exit of the medium but the one averaged along the Rb cell that we have denoted $\Bar{\xi}$. 
Since we know the evolution of I as function of z, we can compute  the effectivea non-linear refractive index $\Delta n$ at the output plane of the cell by taking intou account the absorption:
\begin{equation}
\begin{split}
    \Bar{\Delta n}(L) & =\frac{1}{L} \int^L_0 n_2I(z) \textrm{d}_z,\\
    & =\frac{n_2I_0}{L} \int^L_0  e^{-\alpha z} \textrm{d}_z,\\
    & =\frac{n_2 I_0}{\alpha L}\times (1-e^{-\alpha L}),\\
    & =n_2I_0\frac{T-1}{\ln(T)}.
\end{split}
\end{equation}
Also, we know that $\Delta n(0) = n_2I_0$ and $\Delta n(L)=n_2I_0 e^{-\alpha L}$. Finally, we get:
\begin{equation}
    \Delta n = \Delta n(L) = \Bar{\Delta n}(L)\frac{T\ln(T)}{T-1}.
\end{equation}

\section{Reconstruction of the 2D velocity map}
From the phase map, we reconstruct the total velocity map, defined by $\boldsymbol{v^{tot}}(\boldsymbol{r})=\boldsymbol{\nabla} \theta(\boldsymbol{r})$. 
To avoid any computation error on the phase we must unwrap the phase along both axis, written under the notation $\theta^{'}_x$ and $\theta^{'}_y$. 
The total velocity is then the combination of the $x$ and $y$ component of each gradient, calculated from the unwrapped phase along the two axis, as shown Fig.\ref{fig:velocity_comp}.

\bibliography{biblio}

\end{document}